\documentclass[11pt]{article}
\usepackage[latin1]{inputenc}
\usepackage[english]{babel}
\usepackage[namelimits]{amsmath}
\usepackage{authblk}
\usepackage{amssymb}
\usepackage{amsmath}
\usepackage{amsthm}

{\tiny}

\begin{document}
\title{{\bf The physical origin of the cosmological constant: continuum limit and the analogy with the Casimir effect}}
\author[1,2,3]{S. Viaggiu\thanks{viaggiu@axp.mat.uniroma2.it and s.viaggiu@unimarconi.it}}
\affil[1]{Dipartimento di Fisica Nucleare, Subnucleare e delle Radiazioni, Universit\'a degli Studi Guglielmo Marconi, Via Plinio 44, I-00193 Rome, Italy.}
\affil[2]{Dipartimento di Matematica, Universit\`a di Roma ``Tor Vergata'', Via della Ricerca Scientifica, 1, I-00133 Roma, Italy.}
\affil[3]{INFN, Sezione di Napoli, Complesso Universitario di Monte S. Angelo, Via Cintia Edificio 6, 80126 Napoli, Italy.}

\date{\today}\maketitle
\begin{abstract}
\noindent 
In this paper we continue the investigations in \cite{1} concerning the origin of the cosmological constant. 
First of all, we generalise the results in \cite{1,2} by considering a continuum approximation for a radiation field in a cosmological background. 
In this way, we clearly show that the bare cosmological constant is obtained with wanishing temperature $T$ and that the 
specific heat $C$ is zero at the decoherence scale $L_D$. Moreover, we address the issue to fix the parameters present in our model. In particular, we push forward the analogy between our expression for ${\Lambda}_L$ at a given proper scale $L$ and the one extrapolated by the Casimir effect. As a consequence, we can fix the decoherence scale $L_D$
at which we have the crossover to classicality  to be of the order of $\sim 10^{-5}$ meters. This implies that the actual observed value of the cosmological constant is fixed (frozen) at this new physical scale. 
\end{abstract}

\section{Introduction}

About the $68\%$ of the universe is composed with dark energy. In the $\Lambda$CDM model the dark energy is represented in terms of the cosmological constant $\Lambda$ with energy density ${\rho}_{\Lambda}=\frac{\Lambda c^2}{8\pi G}$ and equation of state
$p_{\Lambda}=-c^2{\rho}_{\Lambda}$. However, a complete physically viable description of the cosmological constant is still
lacking (see for example \cite{2,3,4,5,6,7,8}). Since the cosmological constant can be represented in terms of an energy momentum 
tensor $T_{\mu\nu}$ given by $T_{\mu\nu}\sim\Lambda g_{\mu\nu}$, it is expected that the cosmological constant is provided by vacuum fluctuations. In the usual picture, the observed cosmological constant $\overline{\Lambda}$ is splitted as 
$\overline{\Lambda}=\Lambda+{\Lambda}_{vac}$, where $\Lambda$ is the non-interacting bare (classical)
cosmological constant and ${\Lambda}_{vac}$ represents the contribution due to quantum fluctuations. After introducing a Planckian cutoff $L_P$ and performing naive arguments of quantum field theory, we should have 
$\overline{\Lambda}\sim\frac{1}{L_P^2}$, leading to a value of about $122$ order greater than the one effectively observed. This has been named by Leonard Susskind the wrongest prediction in the hystory of physics. Magic cancellations are often invoked
by the supersymmetry, but this beautiful mechanism has not been observed in the nature. New propals are thus urgent.
To this purpose, in the approach present in \cite{9} the authors use a semi-classical approximation where quantum fluctuations generate a stochastic inhomogeneous metric.
Unfortunately, the model is plagued by several issues, first of all the regularization used is not Lorentz-invariant. This approach has been corrected in \cite{10}, but the price to pay is the introduction of a physically problematic 
micro-cyclic universe with a more problematic 'micro' big bang singularity. Another approach is present in \cite{11}, 
where a dynamical cosmological constant is realised in terms of the Ashtekar variables. Moreover, in \cite{12} the author
depict a scenario where Planckian fluctuations, thanks to the strong inhomogeneities generated on the spacetime,  inhibit a huge cosmological constant created at Planckian scales. Only at macroscopic scales a residual small cosmological constant emerges.\\
Finally, in my proposal \cite{1} a completely new idea is present, There, according with the finding in \cite{12}, 
a very huge positive cosmological constant is created at Planckian scales, but the effective cosmological 
constant depends on the scale at which the physics is considered. As a consequence, the effective value 
of $\overline{\Lambda}$ must be averaged on bigger and bigger scales, up to a decoherence scale $L_D$ where the crossover to classicality emerges. 
In this paper we further analyse the model in \cite{1}. First of all, we study the continuum limit of the radiation field. Moreover, we outline an analogy between our expression for the scale dependent cosmological constant and the one provided by the Casimir effect. With this analogy we can fix the parameters of the model and thus calculate the expected decoherence scale $L_D$.\\
In section 2 we perform the continuum limit. In section 3 we study the cosmological case, while in section 4 we discuss the analogy with the Casimir effect. Finally, section 5 is devoted to some conclusions and final remarks.

\section{Continuum limit for a general equation of state}

In \cite{13,14} we have advanced a physical mechanism, mimicking solid state physics, capable, thanks to Planckian fluctuations, to change the equation of state of a radiation field into an effective one with a $\gamma-$ linear equation of state.
This has been firstly applied in order to explain the logarithmic corrections \cite{13} to the semi-classical Bekenstein-Hawking
entropy \cite{15,16} and further applied in \cite{1} in order to depict the cosmological constant.\\
To start with, we consider the free energy $F^{(0)}(V,N,T)$ of a radiation field with $N$ excitations in a proper volume $V$ at the temperature $T$ and with energy $E^{(0)}$ in the continuum limit, given, as well known, by:
\begin{eqnarray}
& & F^{(0)}=\frac{TV}{c^3\pi^2}\int_{0}^{\infty}\omega^2
\ln\left(1-e^{-\frac{\hbar\omega}{k_B T}}\right)d\omega=-\frac{4\sigma}{3c}V T^4,\label{1}\\
& &\sigma=\frac{\pi^2 k_B^4}{60 c^2\hbar^3},\;\;P^{(0)}V=\frac{E^{(0)}}{3},\;\;E^{(0)}=-3F^{(0)}.\nonumber
\end{eqnarray}
In this framework, we can generalize the results in \cite{1}. Consider the partition function $Z^{(0)}$ of the aforementioned radiation field together with quantum fluctuations. Mimicking solid state physics, we can represent such fluctuations in terms of a field $\Phi(V,N)$ depending on the proper volume $V$, the excitations $N$ and without and explicit dependence on the 
temperature $T$. For the free energy $F$ and the energy $E$ of the 'dressed' system (by fluctuations) we have
\begin{eqnarray}
& &F=-k_B T\ln(Z)=F^{(0)}+\hbar\Phi(N,V),\label{2}\\
& &Z=e^{-\beta\hbar\Phi}Z^{(0)},\;\;E=E^{(0)}+\hbar\Phi(N,V).\label{2a}
\end{eqnarray}
The following proposition holds:\\
{\bf Proposition:} {\it Consider $N$ 
massless excitations within a volume $V$ and a generic regular function $K(E)$ of the energy $E$. The excitations with total energy
$E$ have a generic equation of state
$PV=K(E)$ provided that the differentiable function $\Phi(N,V)$
satisfies the following equation} 
\begin{equation}
\hbar\left[3V\;{\Phi}_{,V}+\Phi\right]=E-3K(E), \label{3}
\end{equation}
{\it together with the condition}
\begin{equation}
E-\hbar\;\Phi > 0.
\label{4}
\end{equation}
\begin{proof}	
Since  we have $E^{(0)}>0$ (radiation field), condition (\ref{4}) follows. 
From (\ref{2}) we have
\begin{eqnarray}
& & F_{,V}=F^{(0)}_{,V}+\hbar{\Phi}_{,V}=-P,\label{5}\\
& & PV=P^{(0)}V-\hbar V{\Phi}_{,V}=\frac{E^{(0)}}{3}-\hbar V{\Phi}_{,V}.\label{6}
\end{eqnarray}
From the expression for $E$ in (\ref{2a}) and after posing $PV=K(E)$ we obtain the (\ref{3}). 
\end{proof}
\noindent The $\gamma-$linear equation of 
state is obtained with $K(E)=\gamma E$. Moreover, in the spherical case with a sphere of proper areal radius $L$ 
we have $V=4\pi L^3/3$ and the equation \cite{1}
\begin{equation}
\hbar\left[L\;{\Phi}_{,L}+\Phi\right]=E(L)(1-3\gamma) \label{7}
\end{equation}
is regained. As an useful example, consider a system with constant energy density $\rho$ and rest energy $E$
given by $E=\rho c^2 V$ and with $K(E)=\gamma E$. 
After integrating the (\ref{3}) and by using the (\ref{1}) we obtain
\begin{equation}
\rho\left[1-\frac{(1-3\gamma)}{4}\right]=\frac{4\sigma}{c^3} T^4. \label{8}
\end{equation}
Existence condition for (\ref{8}) requires that $1-\frac{(1-3\gamma)}{4}>0\rightarrow \gamma >-1$. Hence, according to the results
in \cite{1}, the case $\gamma=-1$ is possible only with $T=0$. This is what we expect from a classical description of the bare cosmological constant $\Lambda$ that is not dressed by quantum fluctuations, i.e. without the
term ${\Lambda}_{vac}$. Also note that the phantom case with $\gamma<-1$ is forbidden in a classical background. 

\section{The cosmological case: temperature and the decoherence scale} 

In this section we apply the results of section above to the cosmological case. 
Without loss of generality, as explained in \cite{1}, we consider a classical de Sitter background given by 
\begin{equation}
ds^2=-c^2dt^2+a^2(t)\left[dr^2+r^2d\theta^2+r^2\sin^2\theta\;d\phi^2\right].
\label{9}	
\end{equation}
A classical Friedmann spacetime with spatial curvature $k$ and Hubble flow $H$
is equipped with a quasi-local energy provided by the Misner-Sharp one $E_{ms}$.
After defining $L=a(t)r$, we have \cite{1}: 
\begin{equation}
E_{ms}=\frac{c^4}{2G}\frac{L^3}{L_A^2},\;\;\;L_A=\frac{c}{\sqrt{H^2+\frac{k}{a^2(t)}}}.
\label{10}
\end{equation}
In the de Sitter case formulas (\ref{10}) hold with $k=0$, $H=c\sqrt\frac{\overline{\Lambda}}{3}$
and $a(t)\sim e^{ct\sqrt\frac{\overline{\Lambda}}{3}}$. In this classical background the formula (\ref{8}) holds and
as a consequence the cosmological constant case with $\gamma=-1$ can be obtained only with $T=0$. This physically means that, without considering quantum fluctuations in some way in the expression for the classical Misner-Sharp energy $E_{ms}$
given by (\ref{10}),
we can depict the classical bare non interacting cosmological constant $\Lambda$ to which it is associated a 
vanishing temperature ($T=0$). To this purpose, note that in a Friedmann context the quasi-local Misner-Sharp energy 
$E_{ms}(L)$ in a given volume of proper areal radius $L$ is nothing but $E_{ms}(L)=\rho c^2 V(L)$, 
as supposed in the formula (\ref{8}).\\
In \cite{1} Planckian fluctuations have been obtained in a suitable way by considering a quantum non-commutative spacetime
\cite{17,18} in a Friedmann flat background \cite{18}. In particular, in \cite{19} physically motivated 
spacetime uncertainty relations (STUR) among the coordinates have been obtained. 
These STUR\footnote{see \cite{1} for more technical details} imply the following expression for the 
quasi-local Misner-Sharp energy (\ref{10}) modified by Planckian fluctuations in a spherical region of proper areal radius
$L$:
\begin{equation}
E(L)=\frac{c^4}{2G}\frac{L^3}{L_A^2}+\xi\frac{c^4}{2G}\frac{L_P^2}{L},
\label{11}
\end{equation}
where $\xi$ is a constant with $\xi\in(0,a),\;a\sim 1$ and $L_A^2=\frac{3}{\Lambda}$. The first term in the right side
of (\ref{11}) is the classical term of the Misner-Sharp energy in terms of the 
bare non-interacting cosmological constant, while the other one represents the correction due to quantum
fluctuations. The constant $\xi$ is a free parameter and its value will be determined in the next section.
Thanks to the expression (\ref{11}) end with the help of  $\overline{\Lambda}=\Lambda+{\Lambda}_{vac}$, for the effective mean density
${\rho}_{\overline{\Lambda}}$ at the scale $L$ we have
\begin{equation}
{\rho}_{{\overline{\Lambda}}_L}=\frac{c^2\Lambda}{8\pi G}+\frac{3\xi c^2}{8\pi G}\frac{L_P^2}{L^4},
\label{12}
\end{equation}
that in turn implies
\begin{equation}
{\overline{\Lambda}}_L=\Lambda+\frac{3\xi L_P^2}{L^4}.
\label{13}
\end{equation}
Note that the expression (\ref{13}) represents the effective cosmological constant at the scale $L$
in terms of bare one $\Lambda$ with added the term proportional to $\xi$ due to Planckian fluctuations. 
At this point of the treatment it is thus necessary to explain the meaning of the dependence on $L$ of (\ref{13}). 
At first look, the behavior (\ref{13}) may seem the one of a dynamical dark energy, but this is not the case. 
The semi-classical expression(\ref{13}) is obtained by fixing a given physical proper scale $L$ and thus considering the
fluctuations inside the proper sphere of areal radius $L$. In fact, in order to calculate the averaged effective cosmological constant acting at the scale $L$, 
we have not at our disposal an accepted quantum gravity theory. To circumvent this issue, we adopted a semi-classical model for the fluctuations to obtain the generalized Misner-Sharp energy inside $L$. If we consider a system at a fixed
$L=constant$ proper areal radius, it is equipped with a cosmological constant that, on average, is provided by the (\ref{13}).
Hence, for the scale keep fixed $L$, we have an effective fixed ${\overline{\Lambda}}_{L=constant}$ and as a result we have an 
effective averaged pressure $P_L$ with $c^2{\rho}_{L=constant}=-P_{L=constant}$. In practice, we use within a 
ball of fixed areal radius $L$ a suitable average of the Einstein equations with an effective averaged energy-momentum tensor
$T_{L\mu\nu}$ equipped with a fixed volume averaged  ${\overline{\Lambda}}_L$ that is considered as a constant for the 
chosen volume $V_L$. Also note that we consider proper lengths because they are suitable in a curved context. As an example, when in a curved context we consider the Planck length $L_P$, this must be considered as a proper length. In a Friedmann flat context,
in relation to (\ref{9}), we have $L_P=a(t)\delta r=const. \simeq 10^{-35}\;m$.\\ 
From the reasonings above it is evident that the Buchert technique \cite{20} can be used in our semi-classical-approach.
In fact in \cite{1} a semi-classical model has been obtained in terms of the Buchert scheme \cite{20}
in order to solve the fitting problem in cosmology but translated at Planckian scales. We can thus suppose that Planckian fluctuations take the effective spacetime metric at a semi-classical level inhomogeneous.
We can thus use the Buchert formalism \cite{20}, but modified at Planckian scales or above, and average an inhomogeneous metric at some slice at constant time in a fixed proper volume of areal radius $L$ and thus obtain a template metric at the scale $L$:
\begin{equation}
ds^2=-c^2 d{t}^2
+a_{L}^2(t)\left[dr^2+r^2\left(d\theta^2+\sin^2\theta d\phi^2\right)\right],
\label{14}	
\end{equation}
where  $a_L$ is the scale factor of the template metric (\ref{14}) satisfying the Buchert equations
\footnote{The explicit expression for $a_L$ plays no role in this paper and for more details see \cite{1}.}.
From a physical point of view the cosmological constant is created at Planckian scale $L_P$ with a huge value 
$\sim 1/L_P^2$. However, when one consider the physics at a given scale $L$ above the Planck one, the effective cosmological constant must be averaged at this scale: since Planckian fluctuations are expected to be monotonically decreasing on
bigger and bigger scales, we expect that the dressing term is monotonically decreasing in term of the scale $L$. In practice, as far as the fitting problem at Planckian scales admits a solution, a given region of proper size $L$ expands with a template metric given by (\ref{14}) and with an effective cosmological constant given by (\ref{13}). The important fact noticed in \cite{1} is that the expression (\ref{11}) has an absolute positive minimum at the scale $L_D={\left(\frac{\xi L_P^2}{\Lambda}\right)}^{\frac{1}{4}}$. The length $L_D$ represents the decoherence scale, that is the scale 
such that the observed value of the cosmological constant is obtained:
\begin{equation}
\overline{\Lambda}(L=L_D)=\Lambda+\frac{3\xi L_P^2}{L_D^4}=4\Lambda=4\xi\frac{L_P^2}{L_D^4}.
\label{15}
\end{equation}
The observed value $\overline{\Lambda}$ is frozen in the lowest energy state of (\ref{11}). In a quantum langauge of a semi-classical approximation, this means that there exists a lowest energy state $\{S_{L_D}\}$ such that the mean value of the component $T_{00}$ of the
energy momentum tensor $T_{\mu\nu}$ is frozen: ${<T_{00}>}_{S_{L_D}}=c^2\overline{\Lambda}$. Modes propagating 
with $L>L_D$ are frozen, the scale $L_D$ thus representing the crossover to classicality. The energy density 
${\rho}_{\overline{\Lambda}}$ for $L>L_D$
is thus frozen at the value it assumes at $L=L_D$.
At $L=L_D$ we have the continuity equation
$\frac{c^4 L_D^3}{2G L_{\overline{\Lambda}}^2}=\frac{c^4}{2G}\frac{L_D^3}{L_{A}^2}+\xi\frac{c^4}{2G}\frac{L_P^2}{L_D}$
with $L_{\overline{\Lambda}}^2=\frac{3}{\overline{\Lambda}}$ and thus for $L>L_D$ the classical expression for the
Misner-Sharp energy emerges. It is worth to be noticed that the resonings above do not imply that $L\sim L_D$ is of the order of the Planck scale, i.e. $L_D$ is not necessarily a scale where the spacetime is quantistic. The decoherence scale can be well above the Planck one, where the manifold is classical and Planckian fluctuations are very small. The scale $L_D$
simply determines the value of the observed cosmological constant.\\
Our model is in some sense phenomenological since it depends on three constant 
$\{\overline{\Lambda},\xi, L_D\}$ that cannot be fixed from the onset. However, formula (\ref{15}) furnishes an intriguing relation between these three parameters. Since $\overline{\Lambda}$ is fixed by astrophysical data to be
$\overline{\Lambda}\sim 10^{-52}/m^2$, this left only two constant, $\{\xi,L_D\}$ unspecified. In the next section we face this problem by a comparison with the Casimir effect.\\
We are now in the position to study our model from a thermodynamical point of view. In particular, we study the scale dependence of the temperature $T$. To start with, we must integrate equation (\ref{7}) with $\gamma=-1$:
\begin{equation}
\hbar\Phi(L)=\frac{c^4}{2G}\frac{L^3}{L_A^2}+\frac{2\xi c^4}{G}
\frac{L_P^2}{L}\ln\left(\frac{L}{L_0}\right).
\label{16}
\end{equation}
The constant $L_0$ in (\ref{16}) is a new integration constant. The condition (\ref{4}) with $E$ given by 
(\ref{11}) is given by $L<L_0 e^{\frac{1}{4}}$. The constant $L_0$ can be fixed in following way. In \cite{21}
it has been shown that the temperature for the cosmological constant in a de Sitter universe is fixed at the apparent horizon 
$L_{\overline{\Lambda}}$ and is given by $T_h=\frac{c\hbar}{4\pi k_B L_{\overline{\Lambda}} }$. It is thus natural to suppose
that at the decoherence scale $L_D$ not only is fixed the value of $\overline{\Lambda}$ but also its temperature.
Hence we can write $T_h=T(L=L_D)=T_D$. By the formula $E=E^{(0)}+\hbar\Phi$, with the help of (\ref{1}),
(\ref{11}) and (\ref{16}) evaluated at $L=L_D$ we obtain:
\begin{equation}
\ln\left(\frac{L_D}{L_0}\right)=\frac{1}{4}-\frac{32\pi G\sigma T_D^4}{3 c^5\overline{\Lambda}}=\frac{1}{4}-\epsilon.
\label{17}
\end{equation}
From (\ref{17}) we have $L_0=L_D e^{-\frac{1}{4}+\epsilon}$ with 
\begin{equation}
\epsilon=\frac{1}{12960\;\pi}\overline{\Lambda} L_P^2.
\label{18}
\end{equation}
Note that the value of $\epsilon$ is extremely small, of the order of $10^{-126}$. This is the adimensional coupling constant
that is expected in a quantum gravity regime in a de Sitter universe from a dimensional point of view. In the limit
$L_P\rightarrow 0$ we have that Planckian contributions are neglected and the term  $\frac{c^4}{2G}\frac{L^3}{L_A^2}$ only is obtained. 
Since  $L<L_0\;e^{\frac{1}{4}}$ we obtain that $L<L_D\;e^{\epsilon}$: this estimation is in agreement with our interpretation
of $L_D$.\\
With the same estimation for (\ref{17}) but evaluated at a generic $L$, for the effective temperature $T_L$ we get:
\begin{equation}
T_L={\left(\frac{3\xi c^5}{32\pi\sigma G}\right)}^{\frac{1}{4}}\frac{\sqrt{L_P}}{L}
{\left[1-4\ln\left(\frac{L}{L_0}\right)\right]}^{\frac{1}{4}},
\label{19}
\end{equation}
with the existence condition $L<L_D\;e^{\epsilon}$. At $L=L_D$ we recover the equation (\ref{17}).
It is also interesting to note that, thanks to the geometrical constraint due to generale relativity, the quasi-local energy content
of a given spherical region in a scale below 
$L_D$ for a Friedmann universe is fixed by the generalized Misner-Sharp energy (\ref{11}) and is a function of the proper areal radius $L$. In turn, the mean temperature $T_L$ of the aforementioned region must be,
thanks to the expression (\ref{2a})  for $E$, a function of $L$. In practice, the quasi-local energy of a given spherical region in a Friedmann spacetime, and thus also of a de Sitter one, it is not arbitrary but fixed by the spacetime geometry
\footnote{Obviously we refer to the physical description of an unperturbed Friedmann spacetime. Nothing forbids us to 
pumping energy in a given region: in this way we perturbate the initial Friedmann geometry, but we are interested in a 
physical decription of the unperturbed de Sitter spacetime and its quasi-local energy is dictated by general relativity.}.
This geometrical constraint is not present in ordinary thermodynamics.\\
Also note that from the (\ref{19}) we deduce that 
$T=0$ happens at $L=L_0 e^{\frac{1}{4}}$ and thus at $L=L_D e^{\epsilon}>L_D$. As stated by equation (\ref{18}) and reasonings below equation (\ref{18}), at $L=L_D$ we have $T_D=T_h\neq 0$ and
consequently $T_L\neq 0\;\forall L\in [L_P,L_D]$. Moreover, the small value of $T_D$ is a consequence
of the smallness of $\epsilon$. Hence, our study suggests that the very small value of $T_h$ for a de Sitter expanding
universe is a consequence of 'residual' Planckian physics at the decoherence scale.\\
Thanks to (\ref{19}) we can characterize the decoherence scale also from a thermodynamical point of view. At $L=L_D$
we have, as stated above, a minimum for $E(L)$ given by (\ref{11}). At this minimum 
we have the crossover to classicality with a temperature given by the horizon one \cite{21} $T_h$ and with the observed value 
$\overline{\Lambda}$. Thanks to (\ref{19}), we can define, as advanced in \cite{1}, a specific heat $C_L$ defined as
$C_L=\frac{dE(L)}{dT}=\frac{dE(L)}{dL}\frac{dL}{dT}$. We have
\begin{equation}
C_L=\frac{c^4}{2G}\left(L^2\frac{\overline{\Lambda}}{4}-\xi\frac{L_P^2}{L^2}\right){\left(\frac{dT}{dL}\right)}^{-1}.
\label{20}
\end{equation}
It is also interesting to study the sign of $C_L$.
To this purpose, after posing 
$Q=\sqrt{L_P}{\left(\frac{3\xi c^5}{32\pi\sigma G}\right)}^{\frac{1}{4}}$ we have
\begin{equation}
\frac{dT}{dL}=-\frac{Q}{L^2}{\left[1-4\ln\left(\frac{L}{L_0}\right)\right]}^{\frac{1}{4}}-
\frac{Q}{L^2}\frac{1}{{\left[1-4\ln\left(\frac{L}{L_0}\right)\right]}^{\frac{3}{4}}}.
\label{21} 
\end{equation}
From the (\ref{21}) we see that $\frac{dT}{dL}<0$ in its domain. As a consequence, from the (\ref{20}) we
deduce that for $L<L_D$ we have $C_L>0$. Note that exactly at $L=L_D$ we have $C_{L_D}=0$. Physically, this means that at the physical scale $L_D$ the system is thermodynamically dead and with frozen termalized temperature $T_h$. Hence,
below the decoherence scale specific heat above defined is strictly positive. This means that, according to physical intuition, as far as the decoherence scale is not reached, the specific heat is positive and zero at the crossover
to classicality. Remember that typically, in a classical self-gravitating system where long range forces are dominant,
the specific heat is expected to be negative. As an extreme example, as well known, the specific heat of a black hole 
is negative. The fact that $C_L$ is positive for $L<L_D$ is an indication that the complete crossover to classicality 
has not yet happened.

\section{A comparison with the Casimir effect: fixing the parameters of our model}

In this section we outline a simple strategy in order to fix the parameters $\{\xi, L_D\}$ of our model. The strategy 
requires a simple analogy with the Casimir effect \cite{22}. To start with, note that in a non-static context, the 
dynamical Casimir effect should be addressed (see for example \cite{k1,k2,k3,k4,k5}). There, a complication arises due to 
particles creation (photons) induced by moving mirrors. However, note that in our model, as pointed in section 3, in order to average Planckian fluctuations in a given proper volume $V(L)$, we fix a physical scale by means of a ball of constant proper areal radius $L$. Hence, at least in a first approximation, we expect that with a geometry such that the 
$L$ is held fixed, photons creation can be neglected\footnote{As an example, in \cite{k2} it has been shown that, within a perturbative approach,dynamical force is the Casimir one modified by a small dynamical
correction.} and as a consequence we are legitimate to use formulas suitable for the static Casimir effect. Moreover, note that, in order to explain the nature of the cosmological constant in (\ref{9}), the scale factor $a(t)$ it is not a function to be determined as in \cite{k2}, but it is fixed by general relativity above the decoherence scale $L_D$ to be
$a(t)\sim e^{ct\sqrt\frac{\Lambda}{3}}$. Hence, it is reasonable to suppose that, thanks to quantum fluctuations and to the physical mechanism depicted by proposition of section 2, massless photons will be transformed into excitations with the equation of state suitable for the cosmological constant, i.e. $K(E)=E,\;\gamma =-1$ and as a result photons eventually created 
simply contribute to the observed value of $\overline{\Lambda}$. These facts can be certainly matter of further investigations.\\
As well known, the experimentally 
verified static Casimir effect \cite{22} establishes the existence of an attractive force between two pefectly conducting
uncharged parallel plates of area $A$ placed in the vacuum at a given distance $d$. The expressions for the force $F$ and the energy $E_{C}$ are:
\begin{equation}
F=-\frac{\hbar\pi^2 c A}{240 d^4},\;\;\;E_C=-\frac{\hbar\pi^2 c A}{720 d^3}.
\label{22}
\end{equation}
The main description of this important effect (see also \cite{23} and references therein) is in terms of quantum vacuum fluctuations represented the zero-point energy of a quantized field. In particular,
since the vacuum fluctuations in the region outside the plates are different with respect to the ones in the inner region, 
vacuum fluctuations exert different forces on the plates with a net effect given by (\ref{22}). This difference is remarkable 
finite despite the fact that the two contributions in the inner and outer region are infinite.
Stated in other words,
the zero point energy is modified due to presence of the plates. From this point of view the Casimir effect can be conceived as a geometrical one depending on the geometry of a given configuration \cite{23}. In this regard, the 
Casimir energy can be seen as the difference between the vacuum energy in presence and in absence of the plates.\\
In order to take contact with our model we must consider the Casimir effect with a different geometry with respect 
to the one initially considered in \cite{22}. In a general context we can interpret the Casimir energy $E_C$ as the difference between vacuum energy in presence of boundary conditions $E_B$ and the one of a space $E_0$ without boundary conditions:
\begin{equation}
E_C=E_B-E_0.
\label{23}
\end{equation}
To go further, we must obtain a suitable formula for the Casimir energy in the case of a spherical region of proper areal radius $L$, where \cite{23} the Casimir energy becomes positive. To this purpose,
in our context, we are tempted to use the expression for $E_C$ in (\ref{22}) with $A(L)\sim L^2$. Hence, we can write:
\begin{equation}
E_C=g\frac{c\hbar}{L},
\label{24}
\end{equation}
where the constant $g$ depends on the geometry of the system. In the case of the Casimir effect for a sphere, we can quote the paper \cite{23}. There, the Casimir energy is positive with a positive repulsive force with $g\simeq 0.046361$. We thus obtain:
\begin{equation}
E_C=\hbar\frac{c\pi^3}{720 L}
\label{25}
\end{equation}
with $g=\frac{\pi^3}{720}\simeq 0.043064$. The next step is to obtain, in light of the formulas (\ref{23}), a suitable expression for $E_C$ in our model. To this regard, the term $E_B$ in our model is given by $E(L)$ given by
(\ref{11}): $E_B\rightarrow E(L)$. This is in fact the expression for the generalized Misner-Sharp energy dressed by
vacuum fluctuations after considering a given physical scale $L$. As stated in section above, a non vanishing temperature given by (\ref{19}) is associated to $E(L)$. The term $E_0$ in (\ref{23}) can be associated to the bare one, that is the one
thai is not dressed by quantum fluctuations. In formulas, we can translate expression (\ref{23}) in the following way: 
\begin{equation}
E_C=E_L(dressed)-E_L(bare).
\label{25a}
\end{equation}
In our context equation (\ref{25a}) is nothing else but
\begin{equation}
E_C=E(L,\overline{\Lambda})-E(L,\Lambda):
\label{26}
\end{equation}  
from (\ref{11}) we thus have
\begin{equation}
E(L)=\frac{c^4\Lambda}{2G}\frac{L^3}{3}+\frac{\xi c^4}{2G}\frac{L_P^2}{L},
\label{27}
\end{equation}
and as a result
\begin{equation}
E_C=E(L,\overline{\Lambda})-E(L,\Lambda)=\frac{\xi c^4}{2G}\frac{L_P^2}{L}.
\label{28}
\end{equation}
It should be noted that in the first calculation in \cite{22} the temperature is supposed to be, in an idealized  calculation, zero. Otherwise, thermal fluctuations due to the presence of dissipation must be taken into account. 
To this purpose, note that in our case a non-vanishing temperature is not due to the presence of a matter field but rather to Planckian fluctuations that in turn create a cosmological constant. 
After equating (\ref{25}) and (\ref{28}) we finally obtain
\begin{equation}
\xi=\frac{\pi^3}{360},
\label{29}
\end{equation}
and the (\ref{15}) becomes
\begin{equation}
\overline{\Lambda}=\frac{\pi^3}{90}\frac{L_P^2}{L_D^4}.
\label{30}
\end{equation}
In our semi-classical (phenomenological as named in \cite{1}) model, we have only one independent parameter. Since we have not yet a way to calculate $L_D$, we can determine its value by quoting $\overline{\Lambda}$ that is given by astrophysical data with $\overline{\Lambda}\simeq\frac{1.8\times 10^{-52}}{m^2}$. We thus obtain the estimation 
$L_D\simeq 10^{-2}$ millimeters. This is the scale at wich is fixed the observed value of the cosmological constant,
representing an absolute minimum for (\ref{11}).\\
Note that, thanks to formula (\ref{15}), the term due to Planckian fluctuations in (\ref{11}) is dominant, as expected, 
with respect to the to classical bar one, for $L<<L_D$. Only approaching the decoherence scale $L=L_D$
the two terms become of the same order of size.\\
As a final interesting consideration, it should be ramarked that in the usual treatment of the Casimir efffect, 
both terms $E_B$ and $E_0$ in (\ref{23}) are diverging, but with the difference finite. In our case both terms
$E(dressed)$ and $E(bare)$ are finite. This is because the presence of a quantum spacetime at Planckian scales naturally introduce a cutoff for the shape of a spherical region with $L\geq L_P$. Hence in our model an ultraviolet cutoff is 
provided by the presence of a quantum spacetime at Planckian lengths.

\section{Conclusions and final remarks}
In this paper we continued the investigations in \cite{1}. In this regard the observed cosmological constant 
$\overline{\Lambda}$ is shifted in the usual way $\overline{\Lambda}=\Lambda+{\Lambda}_{vac}$. The term $\Lambda$
represents the bare cosmological constant, that can be obtained at $T=0$ thanks to the calculations leading to formula (\ref{19}), while the term
${\Lambda}_{vac}$ is the contribution due to fluctuations. This shift is translated into the expression
(\ref{11}) representing a generalization of the quasi-local Misner-Sharp energy dressed by Planckian fluctuations. The important 
concept of our model is the introduction of a decoherence scale, representing an absolute minimum for $E(L)$, and fixing the observed value for the cosmological constant $\overline{\Lambda}$. The fundamental ingredient represented by $L_D$ solves the apparent paradox in \cite{24}. In fact, if the scale fixing $\overline{\Lambda}$ is the one provided by the apparent horizon, the expected $\overline{\Lambda}$ will be too small. The presence of $L_D$ of the order of $10^{-5}$ meters, as estimated in this paper, it gives a reasonable solution to the issue in \cite{24}, without introducing the hypothesis
that vacuum fluctuations act also at cosmological scales.\\
It is important to stress that the effective cosmological constant in (\ref{13}) has been obtained by a suitable average procedure after fixing a given proper volume $V_L$. Hence, the effective value of the averaged cosmological constant
$\Lambda_L$ is fixed once a certain proper volune average is chosen. As a result, for the averaged Einstein equations using the
Buchert scheme \cite{20}, the equation of state $c^2{\rho}_L=-P_L$ is satisfied.\\
In the second part of this paper we studied the analogy between our formula (\ref{11}) and the well known Casimir effect, in particular with the calculations in \cite{23} for a spherical conductor. In fact, in the usual treatment of vacuum fluctuations
the following expression for the energy is used
\begin{equation}
E=\frac{4\sigma}{c}V T^4+\sum_{modes}^{\omega_{max}}\frac{\hbar\omega}{2}.
\label{31}
\end{equation}
The UV cutoff $\omega_{max}$ is taken of Planckian size, leading to the well known cosmological constant catastrophe.
In our model, formula (\ref{31}) is substituted, at a given scale $L$, with
\begin{eqnarray}
& &E(L)=\hbar\Phi(L)+\frac{4\sigma}{c}V T^4=\label{32}\\
& &=\frac{4\sigma}{c}V T^4+\frac{c^4}{2G}\frac{L^3}{L_A^2}+\frac{2\xi c^4}{G}
\frac{L_P^2}{L}\ln\left(\frac{L}{L_0}\right). \label{33}
\end{eqnarray}
The last two terms in (\ref{33}) represent the finite contribution of Planckian fluctuations, motivated by the results in
\cite{19}, with the first one denoting the bare component. The further ingredient provided by general relativity 
refers to the expression of $E(L)$. As well known, energy is not local in general  relativity. The closest concept to a local energy in general relativity in a cosmological context is provided by the Misner-Sharp mass. It is thus physically reasonable to assume for 
$E(L)$ the quai-local Misner-Sharp energy (\ref{11}) corrected by Planckian fluctuations. 
In order to realize a comparison with the Casimir effect, we have used formula (\ref{26}), that in turns is motivated by the
(\ref{23}). As a result, we have obtained formulas (\ref{29}) and (\ref{30}) that permit us to calculate the value for the decoherence scale $L_D$, since $\overline{\Lambda}$ is known from astrophysical data.\\
Note that the subtraction procedure (\ref{25a}) eliminates the bare 'non-local term $\frac{c^4\Lambda}{2G}\frac{L^3}{3}$ leaving only the 'local' dressing one $\frac{\xi c^4}{2G}\frac{L_P^2}{L}$.\\ 
The model presented in \cite{1} and further developed in this paper furnishes a possible new way to treat Planckian fluctuations. In fact, we take seriously the idea that Planckian fluctuations are created at Planckian scales, but when one considers a certain 
spherical region above the Planck scale, fluctuations must be averaged on in order to obtain an effective expression for the energy-density and the cosmological constant. We obtain a behavior for quantum modification of the Misner-Sharp mass that is in agreement with the Casimir energy $E_C$ obtained in the standard Casimir effect for a spherical conductor together with a reasonable value for $\xi$.\\
As a final consideration, note that the ratio $\Delta$ between the dressed and the bare terms in (\ref{11}), thanks
to (\ref{15}), is given by $\Delta=3\frac{L_D^4}{L^4}$. At scales $L\sim 10^{-1} L_D\simeq 10^{-6}$ meters, we have
$\Delta\sim 10^{4}$ and as a consequence also at scale at which Casimir experimenst are performed, the second term in (\ref{11})
is still dominant. Only at scales $L\simeq L_D$ $\Delta\simeq 1$: this further justifies the scale $L_D$ as denoting the crossover to classicality.

\end{document}